# Temperature-dependent absolute refractive index measurements of synthetic fused silica

Douglas B. Leviton[*] and Bradley J. Frey
NASA Goddard Space Flight Center, Greenbelt, MD 20771

## ABSTRACT

Using the Cryogenic, High-Accuracy Refraction Measuring System (CHARMS) at NASA's Goddard Space Flight Center, we have measured the absolute refractive index of five specimens taken from a very large boule of Corning 7980 fused silica from temperatures ranging from 30 to 310 K at wavelengths from 0.4 to 2.6 microns with an absolute uncertainty of $\pm 1 \times 10^{-5}$. Statistical variations in derived values of the thermo-optic coefficient (dn/dT) are at the $\pm 2 \times 10^{-8}$/K level. Graphical and tabulated data for absolute refractive index, dispersion, and thermo-optic coefficient are presented for selected wavelengths and temperatures along with estimates of uncertainty in index. Coefficients for temperature-dependent Sellmeier fits of measured refractive index are also presented to allow accurate interpolation of index to other wavelengths and temperatures. We compare our results to those from an independent investigation (which used an interferometric technique for measuring index changes as a function of temperature) whose samples were prepared from the same slugs of material from which our prisms were prepared in support of the Kepler mission. We also compare our results with sparse cryogenic index data from measurements of this material from the literature.

## 1. INTRODUCTION

High quality, refractive optical designs depend intimately on accuracy of refractive index data of constituent optical materials. Since absolute refractive index is generally a function of both wavelength and temperature, it is important to know refractive indices at the optical system's design operating temperature. Further, for large, refractive, optical components, spatial variation of both a material's refractive index and its thermo-optic coefficient (dn/dT) can be potentially detrimental to optical system performance, so spatial knowledge of dn/dT is also important.

The refractive index of fused silica and its dependence on temperature have been studied by a number of investigators using various techniques, both above and below room temperature. In 1965, Malitson reported on the room temperature interspecimen variability in refractive index of optical quality fused silica from three manufacturers using the method of minimum deviation in air from the near ultraviolet to 3.37 microns with a reported error of $\pm 0.5 \times 10^{-5}$ for the visible to $\pm 2 \times 10^{-5}$ in the infrared. He also developed a dispersion relation for fused silica which has been well-trusted since that time.[1] In 1969, Wray and Neu measured refractive index of Corning 7940 synthetic fused silica in vacuum with a reported error of $\pm 2 \times 10^{-4}$ from 300-1100 K from the near ultraviolet to 3.37 microns also using the method of minimum deviation.[2] In 1971, Waxler and Cleek measured changes in refractive index by observing shifts in Fizeau interference fringes with temperature in a plate of fused silica from room temperature to 81 K for 10 visible lines.[3] They calculated refractive index by offsetting room temperature data of Malitson by their measured index changes. While their measurements were made in vacuum, their results are reported in air. In 1991, Matsuoka et al. measured refractive index of Type III silica glass (Nippon Seiki Glass Company) in vacuum with a reported precision of $\pm 3 \times 10^{-6}$ from 108-356 K at 10 lines from the near ultraviolet to the mid-visible, using the method of minimum deviation.[4]

We have conducted a thorough study of the absolute refractive index of Corning 7980 synthetic fused silica by the method of minimum deviation using the Cryogenic High Accuracy Refraction Measuring System (CHARMS) at GSFC.[5,6,7] This paper contains two discussions of the cryogenic refractive index of fused silica based on recent measurements. The first discussion pertains to the dedicated study of five prisms made from core sections cut from around the perimeter of a one meter sized optical blank used for the Schmidt corrector plate for the photometer telescope for NASA's Kepler mission over a wavelength and temperature range applicable to the photometer in flight. Our refractive index and dn/dT data are compared to independent measurements commissioned by the Kepler Project from Precision Measurements and Instruments Corporation (PMIC) in Corvallis, Oregon, of twin specimens taken from those same core sections. The second discussion documents a more general study of the material from the visible out to the strong absorption feature at 2.75 microns in the infrared from room temperature down to about 30 K – the lowest temperature achievable with this material in CHARMS. Our refractive index data, which extend the ranges of wavelength and temperature coverage beyond those of investigations listed in the previous paragraph, will be compared with those from the latter where they overlap.

---
[*] doug.leviton@nasa.gov, phone 1-301-286-3670, FAX 1-301-286-0204

Already, to date, in addition to their use for the Kepler mission, these refractive index values have been employed in the designs of several other cryogenic optical systems for NASA, including the weak lenses for the fine phasing system for the primary mirror segments on the James Webb Space Telescope (JWST), the pupil imaging lenses for the JWST Near Infrared Camera (NIRCam), and the refractive phase plates for the cryogenic optical verification stimulus for NIRCam.

## 2. STUDY FOR KEPLER PHOTOMETER

The Kepler mission is designed to detect the presence of Earth-like planets around other stars by doing extremely precise photometry of stellar systems over time to find variations in their light output which would indicate the presence of orbiting planets. Invariance of the Kepler telescope's point spread function (PSF) at each field position is a crucial performance aspect of the telescope as temporal variations in the PSF might be mistaken for variation in stellar system light output. As such, spatial variations in the refractive index of the telescope's Schmidt corrector, whether intrinsic or induced in some other way such as through a temperature gradient, are intolerable for mission success.

When the roughly 1 m diameter optical blank for the corrector, made of Corning 7980 fused silica, was prepared, five cores 51 mm in diameter and 51 mm thick were also taken from around the perimeter of the blank, evenly spaced in angle every 72°. The Kepler Project became concerned whether variations in refractive index with gradients in temperature in the corrector would compromise the constancy of the telescope's PSF and commissioned measurements of spatial thermo-optic coefficient using the various core samples over the cryogenic operating temperature range of the photometer at both the CHARMS facility and at PMIC. Each core was cut in two and from the two pieces, test specimens were prepared for each facility: a prism for minimum deviation refractometry in CHARMS, and a plate form for interferometric study at PMIC. Each specimen was tagged with its angular position in degrees around the blank using the following serial numbers: 72, 144, 216, 288, and 360.

The prisms used in CHARMS had a nominal apex angle of 59°, refracting face length of 38.1 mm, and height of 28 mm. (The choice of 59° is preferable to 60° only in that when measuring the apex angle using an autocollimator, confusion can come about over which autocollimator return is which with an equilateral or isosceles right prism due to internal reflections. This confusion is avoided with a 59° isosceles prism.) The test specimens for PMIC's interferometric study were 51 mm in diameter, and 25 mm thick polished plates.

The Kepler Photometer's spectral coverage extends from 413 – 914 nm, and its operating temperature range is from -110 to –45 C or 163 to 228 K. At PMIC, only changes in refractive index with temperature are measured using a fringe counting interferometric technique employing a HeNe laser of 0.6328 microns wavelength. PMIC can also measure at 1.064 microns using a Nd:YAG laser, but only data at 0.6328 microns is used in this comparison. Because the physical thickness of a specimen changes with temperature, PMIC actually calculates changes in index, dn, from a combination of fringe counting data and direct measurements of the coefficient of thermal expansion (CTE) of the specimen over the same temperature range. PMIC's measurements covered the temperature range from roughly 155 to 295 K.

In CHARMS, absolute refractive index, n, was directly measured for all five prisms over the temperature range 135 to 305 K at the following wavelengths: 0.4, 0.45, 0.5, 0.55, 0.6, 0.6328, 0.65, 0.7, 0.75, 0.8, 0.85, 0.9, 0.95, 1.0, 1.05, 1.064, 1.10, and 1.15 microns. The laser wavelengths 0.6328 and 1.064 microns were included in the list to allow direct comparison of dn from CHARMS with that from PMIC. Thermo-optic coefficient from CHARMS measurements is simply taken as the derivative of n(T) with respect to temperature. The light source for CHARMS was a quartz tungsten halogen (QTH) lamp feeding a monochromator with a calibrated wavelength accuracy of 0.2 nm. Uncertainty in measured refractive index in CHARMS depends on wavelength and temperature-dependent dispersion and thermo-optic coefficient of the sample material as well as knowledge of wavelength and temperature, respectively. Worst case uncertainties in refractive index for our measurements on fused silica are listed in Table 1 for representative wavelengths and temperatures. Larger values of uncertainty in n beyond a wavelength of 2 microns are due to a dramatic increase in dispersion in the material approaching the deep absorption feature at 2.75 microns. Peak deviation of dn/dT for any given sample compared to the global values of dn/dT averaged over all five samples is $5 \times 10^{-8}$/K over the wavelength and temperature range considered in this study, while noise in dn/dT for a given sample is of the order of $\pm 2 \times 10^{-8}$/K.

Table 1 – uncertainty in measured refractive index of fused silica in CHARMS for representative wavelengths and temperatures.

| wavelength [um] | 30 K | 80 K | 150 K | 250 K | 295 K |
|---|---|---|---|---|---|
| 0.5 | 0.000011 | 0.000011 | 0.000011 | 0.000010 | 0.000010 |
| 1.0 | 0.000008 | 0.000009 | 0.000009 | 0.000008 | 0.000007 |
| 1.5 | 0.000009 | 0.000010 | 0.000010 | 0.000009 | 0.000008 |
| 2.0 | 0.000014 | 0.000015 | 0.000014 | 0.000014 | 0.000014 |
| 2.5 | 0.000017 | 0.000018 | 0.000017 | 0.000017 | 0.000017 |

An initial check of the reasonableness of our measurements of absolute refractive index using CHARMS involves comparing our measured values at room temperature to the well-accepted dispersion law for synthetic fused silica in air of Malitson which is known to be based at least in part on Corning 7940 material.[1] In order to compare our absolute measurements to the dispersion law, we adjust the dispersion law to vacuum assuming the index of air at room temperature to be 1.00027. Our measurements agree with that dispersion law to generally well less than $1 \times 10^{-5}$ and typically to less than $\pm 5 \times 10^{-6}$ until the absorption feature starting at about 2 microns is reached. The departure of the two fits beyond the level of our stated uncertainties past 2.25 microns is explainable through differences in purity of the materials available now and back in 1965 as well as in the way refractometers of different construction treat materials in wavelength regions where materials are absorbing.

In most cases, our in-process check of raw refractive index data from CHARMS involves fitting measured index at each wavelength to a second order polynomial with temperature. This was done for the raw index data for each of the five prisms over the stated temperature range. That the residuals for any such fit for the five samples – $<2 \times 10^{-6}$ rms, $< 2 \times 10^{-6}$ on an absolute average basis, and $+9/-6 \times 10^{-6}$ on a peak-to-valley basis – are so small indicates that variation of index with temperature at a given wavelength is in fact quadratic over the temperature range considered. Figure 1 shows a plot of every index measurement at 633 nm for each of the prisms compared to a global quadratic fit over all measurements (over 5000 of them) for the five prisms. The peak departure of any single measurement from the global fit is $1 \times 10^{-5}$ while the rms departure from the global fit is $4 \times 10^{-6}$. Therefore, we conclude that the samples are indistinguishable in refractive index at any temperature to within our stated uncertainty for a single measurement.

Figure 2 is a plot of PMIC's measurements of changes in index, dn, with temperature for four of the five specimens provided. Specimen 360 was apparently tested in four separate runs, while specimen 72 was temperature cycled several times during one run. Specimens 216 and 288 were temperature cycled once, while we have no data for specimen 144. There is a significant spread of measured dn's across samples and even for the one sample which was cycled numerous times. There also seem to be two knees in each dn(T) curve (one at -70 C and one at -35 C) of unknown origin. These presumably have something to do with the method by which PMIC combines measured fringe counts and CTE data.

Table 2 compares PMIC's measurements of changes in index n with temperature to measurements using CHARMS for the same samples. To compare, we first computed average values of dn with respect to room temperature for all samples using PMIC's fits of dn(T) for each sample for several temperatures in the Kepler Photometer's operating temperature range. We then derived values of n at those temperatures based on PMIC's average dn's assuming the value of n at room temperature from CHARMS was correct. We then took the difference in index, Δn, at each temperature (in parts per million) between those derived values and values from the global fit of CHARMS index data shown in Figure 1.

One can see that PMIC's values depart monotonically from those from CHARMS as temperatures get lower and lower. Meanwhile, one can see that those departures are covered by the spread in PMIC's dn values across samples, which are typically of the order of $5 \times 10^{-5}$ – about an order of magnitude larger than the rms spread in dn values from CHARMS. It is not entirely surprising that PMIC's data show greater spread as PMIC measures two separate physical phenomena and combines them in some fashion into one interpretation for dn.

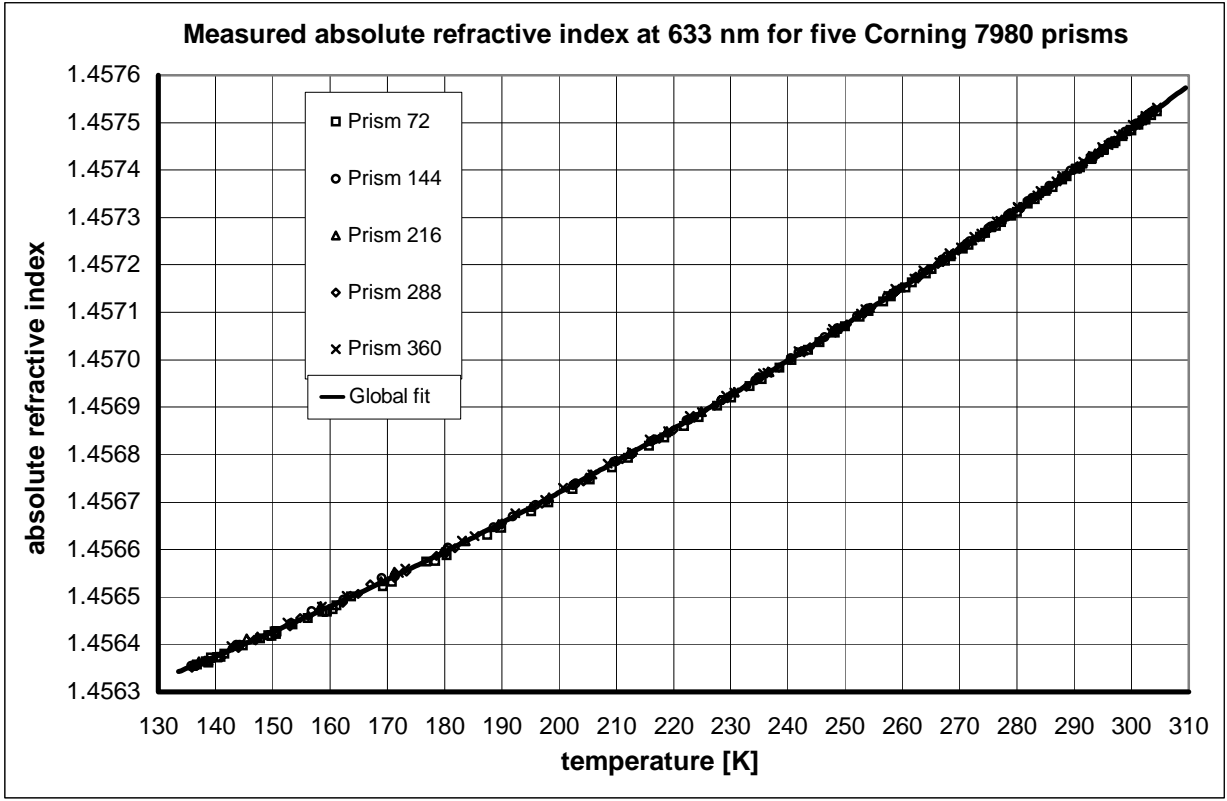

Figure 1 – individual index measurements in CHARMS for five fused silica prisms compared to global fit over all measurements

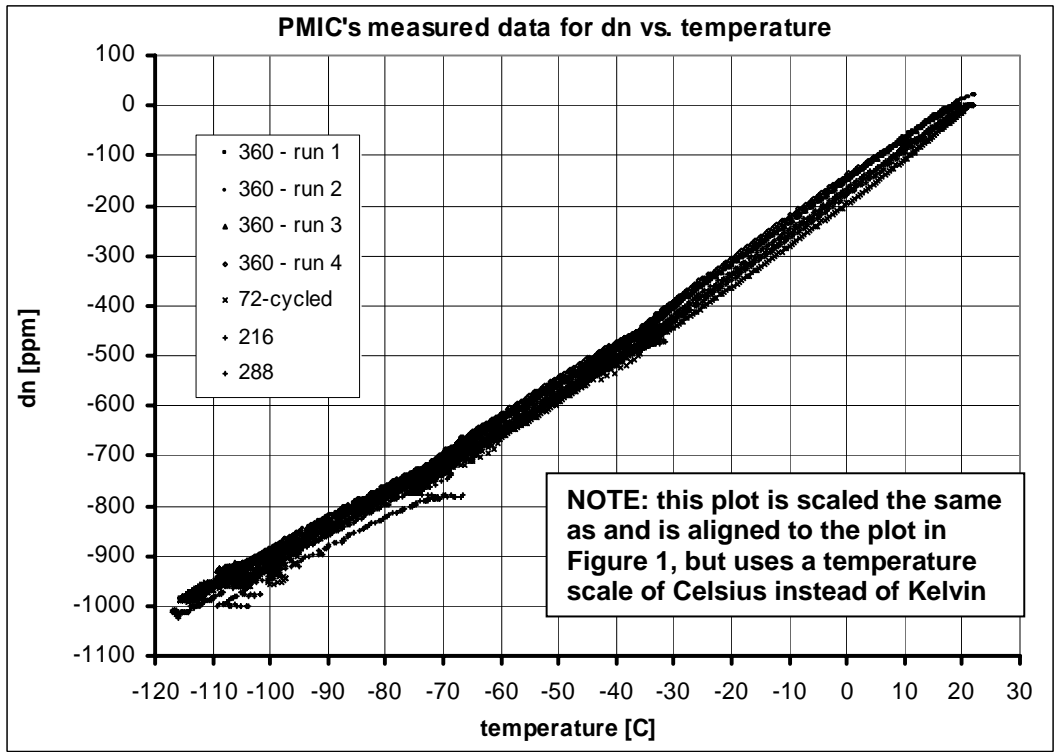

Figure 2 – PMIC's measurements of dn(T) for four fused silica specimens

A comparison of measured thermo-optic coefficient at 633 nm as a function of temperature for PMIC and CHARMS is shown in Figure 3. PMIC's value is everywhere higher than that from CHARMS, but the difference between the two shrinks to lower temperatures as dn/dT itself shrinks. Also shown is the spread of values of dn/dT across samples at each facility. The spread in PMIC's values is smallest around 265 K ($15 \times 10^{-8}$/K) and is larger at lower temperatures (up to $150 \times 10^{-8}$/K). The spread in CHARMS's values is nowhere larger than $9 \times 10^{-8}$/K for this wavelength.

Table 2 – differences in measured dn between PMIC and CHARMS facilities for fused silica
over the operating temperature range for the Kepler Photometer

| Temperature [K] | 160 | 170 | 180 | 190 | 200 | 210 | 220 | 230 | 295 |
|---|---|---|---|---|---|---|---|---|---|
| Δn: PMIC - CHARMS (ppm) | -56 | -53 | -50 | -46 | -42 | -39 | -34 | -30 | 1 |
| spread in dn from PMIC fits [ppm] | 66 | 59 | 53 | 47 | 42 | 37 | 36 | 37 | 34 |

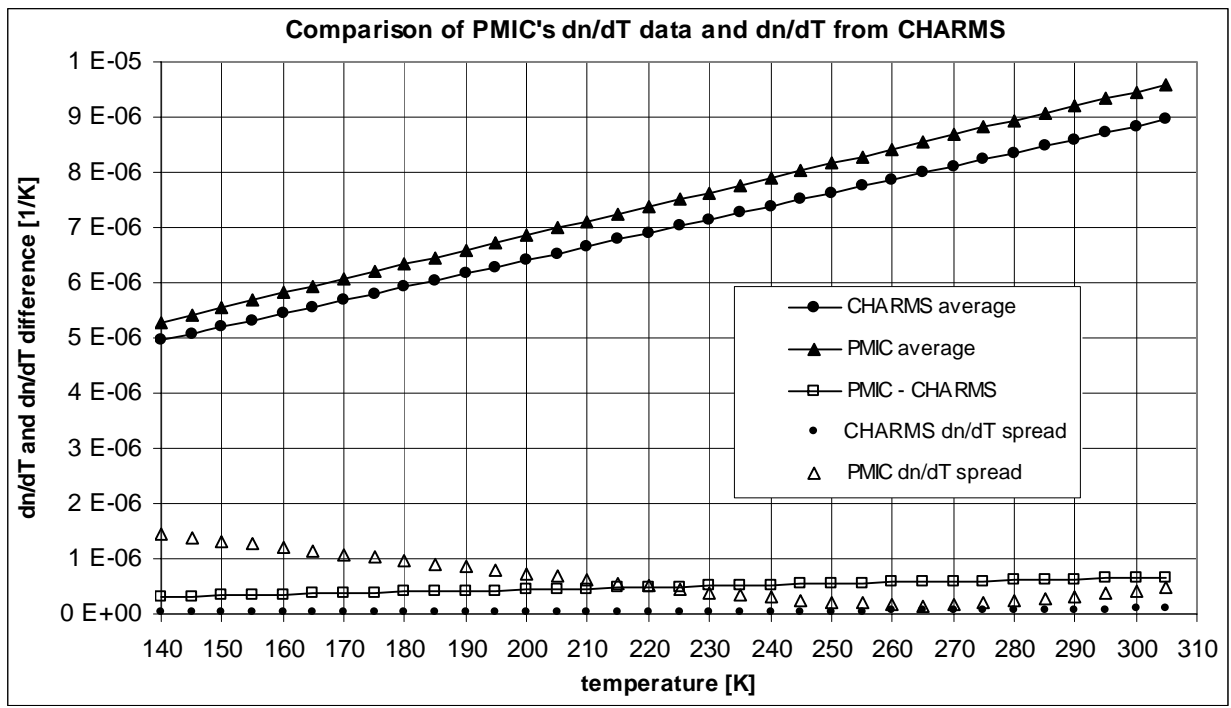

Figure 3 – average thermo-optic coefficient at 633 nm for five fused silica specimens from 140 – 305 K from PMIC and CHARMS

## 3. REFRACTIVE INDEX OF FUSED SILICA FROM 30 - 310 K AND 0.4 – 2.6 MICRONS

All five prisms used in the study were found to be identical in refractive index over the temperature range of interest to the Kepler Photometer to within our ability to tell. Thus, we selected two of the five arbitrarily for study over the expanded wavelength and temperature range accessible to CHARMS. Although fused silica transmits in the infrared between about 2.9 to about 3.75 microns (longward of the absorption feature centered at 2.75 microns), we anticipated few practical applications of the material in that wavelength range and decided to forego measurements in that range to spare the cost of such measurements. Meanwhile, we had previously made some measurements in that range with a single commercial, off-the-shelf fused silica prism we believe to have been Corning 7940 and found its measured indices agreed well with Malitson's dispersion law (corrected to vacuum) in that wavelength range at room temperature.

Once all refractive indices have been measured over the desired wavelength and temperature range, we use a computer program we call the CHARMS Data Cruncher (CDC) to examine raw data from the refractometer and reduce it to the point where resulting measured index values can be fit to a Sellmeier equation. In order to get in-process assessments of data quality during CDC runs, measured index values for each wavelength are fit piecewise to second order polynomials in temperature of the form $n(T) = a \cdot T^2 + b \cdot T + c$ above and below some selected crossover temperature. In order to compute spectral dispersion, first a table of index values is computed on a regular wavelength and temperature grid from these piecewise quadratic fits. From that table, a new table of spectral dispersion, $dn/d\lambda$, is computed by dividing differences in index value, n, by corresponding differences in wavelength, $\lambda$, for each temperature. Thermo-optic coefficient, $dn/dT$, is simply the first derivative of n(T) with respect to T or $dn(T)/dT = 2a \cdot T + b$. CDC produces a table of thermo-optic coefficients on the same regular wavelength and temperature grid described above.

CDC also produces a table of estimated index errors for different wavelength and temperature combinations as seen in Table 1 for fused silica  A partial index error dn is computed for each of four factors (based on presumably known uncertainties in those factors), and the four resulting dn's are combined in quadrature to produce a net index error estimate. The four partial dn's are computed using: 1) worst case uncertainty in calibrated wavelength, $d\lambda$, along with computed $dn/d\lambda$; 2) worst case uncertainty in measured temperature, dT, along with computed $dn/dT$; 3) worst case uncertainty in measured apex angle, $d\alpha$, along with analytically derived $dn/d\alpha$; and 4) worst case uncertainty in measured beam deviation angle, $d\delta$, along with analytically derived $dn/d\delta$.

To obtain a useful dispersion law for the material at hand, we fit raw measured indices for each material to a three term temperature-dependent Sellmeier model of the form:

$$n^2(\lambda,T) - 1 = \sum_{i=1}^{3} \frac{S_i(T) \cdot \lambda^2}{\lambda^2 - \lambda_i^2(T)}$$

where,

$$S_i(T) = \sum_{j=0}^{4} S_{ij} \cdot T^j$$

$$\lambda_i(T) = \sum_{j=0}^{4} \lambda_{ij} \cdot T^j$$

where $S_i$ would be the strengths of the resonance features in the material at wavelengths $\lambda_i$, over all wavelengths and temperatures measured. When dealing with a wavelength interval between wavelengths of physical resonances in the material, the summation may be approximated by typically only three terms. In such an approximation, resonance strengths $S_i$ and wavelengths $\lambda_i$ no longer have direct physical significance but are rather parameters used to generate an adequately accurate fit to empirical data. The Sellmeier model is our best statistical representation of the measured data over the complete measured ranges of wavelength and temperature. The coefficients of the resulting Sellmeier model for Corning 7980 fused silica are tabulated in Table 3.

Table 3 – coefficients for the three term Sellmeier model with 4[th] order temperature dependence for Corning 7980 fused silica

| | Coefficients for the temperature dependent Sellmeier equation for Corning 7980 fused silica | | | | | |
| | 30 K <= T <= 300 K; 0.4 µm <= λ <= 2.6 µm | | | | | |
|---|---|---|---|---|---|---|
| | $S_1$ | $S_2$ | $S_3$ | $\lambda_1$ | $\lambda_2$ | $\lambda_3$ |
| **Constant term** | 1.10127E+00 | 1.78752E-05 | 7.93552E-01 | -8.90600E-02 | 2.97562E-01 | 9.34454E+00 |
| **T term** | -4.94251E-05 | 4.76391E-05 | -1.27815E-03 | 9.08730E-06 | -8.59578E-04 | -7.09788E-03 |
| **$T^2$ term** | 5.27414E-07 | -4.49019E-07 | 1.84595E-05 | -6.53638E-08 | 6.59069E-06 | 1.01968E-04 |
| **$T^3$ term** | -1.59700E-09 | 1.44546E-09 | -9.20275E-08 | 7.77072E-11 | -1.09482E-08 | -5.07660E-07 |
| **$T^4$ term** | 1.75949E-12 | -1.57223E-12 | 1.48829E-10 | 6.84605E-14 | 7.85145E-13 | 8.21348E-10 |

Absolute refractive indices of Corning 7980, based on the three term, temperature-dependent Sellmeier fit described in Table 3, are tabulated in Table 4 and plotted in Figure 4 for selected temperatures and wavelengths.  Spectral dispersion is tabulated in Table 5 and plotted in Figure 5.  Thermo-optic coefficient is tabulated in Table 6 and plotted in Figure 6.

Table 4 – absolute refractive index, n, of Corning 7980 fused silica at selected wavelengths and temperatures

| wavelength | 30 K | 40 K | 50 K | 60 K | 80 K | 100 K | 120 K | 160 K | 200 K | 240 K | 275 K | 295 K | 300 K |
|---|---|---|---|---|---|---|---|---|---|---|---|---|---|
| 0.4 microns | 1.46899 | 1.46902 | 1.46905 | 1.46907 | 1.46912 | 1.46918 | 1.46926 | 1.46948 | 1.46974 | 1.47005 | 1.47036 | 1.47054 | 1.47059 |
| 0.5 microns | 1.46129 | 1.46131 | 1.46133 | 1.46135 | 1.46140 | 1.46146 | 1.46154 | 1.46175 | 1.46199 | 1.46228 | 1.46257 | 1.46275 | 1.46279 |
| 0.6 microns | 1.45704 | 1.45706 | 1.45708 | 1.45710 | 1.45715 | 1.45721 | 1.45729 | 1.45748 | 1.45773 | 1.45801 | 1.45829 | 1.45846 | 1.45850 |
| 0.7 microns | 1.45432 | 1.45433 | 1.45435 | 1.45437 | 1.45442 | 1.45448 | 1.45456 | 1.45475 | 1.45499 | 1.45527 | 1.45554 | 1.45571 | 1.45575 |
| 0.8 microns | 1.45236 | 1.45237 | 1.45239 | 1.45240 | 1.45245 | 1.45252 | 1.45259 | 1.45278 | 1.45302 | 1.45329 | 1.45356 | 1.45373 | 1.45377 |
| 0.9 microns | 1.45080 | 1.45081 | 1.45083 | 1.45085 | 1.45090 | 1.45096 | 1.45104 | 1.45122 | 1.45146 | 1.45173 | 1.45200 | 1.45216 | 1.45220 |
| 1.0 microns | 1.44947 | 1.44948 | 1.44950 | 1.44952 | 1.44957 | 1.44963 | 1.44970 | 1.44989 | 1.45012 | 1.45039 | 1.45066 | 1.45082 | 1.45086 |
| 1.2 microns | 1.44711 | 1.44713 | 1.44714 | 1.44716 | 1.44721 | 1.44727 | 1.44735 | 1.44753 | 1.44776 | 1.44803 | 1.44829 | 1.44845 | 1.44850 |
| 1.5 microns | 1.44369 | 1.44370 | 1.44372 | 1.44374 | 1.44379 | 1.44385 | 1.44392 | 1.44411 | 1.44433 | 1.44460 | 1.44486 | 1.44502 | 1.44507 |
| 1.6 microns | 1.44250 | 1.44251 | 1.44252 | 1.44254 | 1.44259 | 1.44265 | 1.44272 | 1.44291 | 1.44314 | 1.44340 | 1.44367 | 1.44383 | 1.44387 |
| 1.8 microns | 1.43995 | 1.43996 | 1.43998 | 1.44000 | 1.44004 | 1.44010 | 1.44018 | 1.44036 | 1.44059 | 1.44086 | 1.44112 | 1.44128 | 1.44132 |
| 2.0 microns | 1.43716 | 1.43717 | 1.43718 | 1.43720 | 1.43725 | 1.43731 | 1.43738 | 1.43757 | 1.43779 | 1.43806 | 1.43833 | 1.43849 | 1.43853 |
| 2.2 microns | 1.43407 | 1.43408 | 1.43410 | 1.43411 | 1.43416 | 1.43422 | 1.43430 | 1.43448 | 1.43471 | 1.43498 | 1.43524 | 1.43541 | 1.43545 |
| 2.4 microns | 1.43065 | 1.43067 | 1.43068 | 1.43070 | 1.43074 | 1.43081 | 1.43088 | 1.43107 | 1.43129 | 1.43156 | 1.43183 | 1.43200 | 1.43204 |
| 2.5 microns | 1.42881 | 1.42882 | 1.42884 | 1.42885 | 1.42890 | 1.42896 | 1.42904 | 1.42922 | 1.42945 | 1.42972 | 1.42999 | 1.43016 | 1.43021 |
| 2.6 microns | 1.42688 | 1.42688 | 1.42690 | 1.42692 | 1.42696 | 1.42703 | 1.42710 | 1.42729 | 1.42752 | 1.42779 | 1.42806 | 1.42823 | 1.42828 |

Table 5 – spectral dispersion, dn/dλ, in Corning 7980 fused silica at selected wavelengths and temperatures in units of 1/microns

| wavelength | 30 K | 40 K | 50 K | 60 K | 80 K | 100 K | 120 K | 160 K | 200 K | 240 K | 275 K | 295 K | 300 K |
|---|---|---|---|---|---|---|---|---|---|---|---|---|---|
| 0.45 microns | -0.0772 | -0.0772 | -0.0772 | -0.0772 | -0.0772 | -0.0772 | -0.0772 | -0.0774 | -0.0775 | -0.0777 | -0.0779 | -0.0780 | -0.0780 |
| 0.5 microns | -0.0561 | -0.0561 | -0.0561 | -0.0561 | -0.0562 | -0.0562 | -0.0562 | -0.0563 | -0.0564 | -0.0565 | -0.0566 | -0.0567 | -0.0567 |
| 0.6 microns | -0.0338 | -0.0338 | -0.0338 | -0.0338 | -0.0338 | -0.0338 | -0.0338 | -0.0339 | -0.0339 | -0.0340 | -0.0341 | -0.0341 | -0.0341 |
| 0.7 microns | -0.0228 | -0.0228 | -0.0228 | -0.0228 | -0.0228 | -0.0228 | -0.0228 | -0.0228 | -0.0229 | -0.0229 | -0.0229 | -0.0230 | -0.0230 |
| 0.8 microns | -0.0173 | -0.0173 | -0.0174 | -0.0174 | -0.0174 | -0.0174 | -0.0174 | -0.0174 | -0.0174 | -0.0174 | -0.0175 | -0.0175 | -0.0175 |
| 0.9 microns | -0.0142 | -0.0142 | -0.0142 | -0.0142 | -0.0142 | -0.0142 | -0.0142 | -0.0142 | -0.0142 | -0.0142 | -0.0143 | -0.0143 | -0.0143 |
| 1.0 microns | -0.0127 | -0.0127 | -0.0127 | -0.0127 | -0.0127 | -0.0127 | -0.0127 | -0.0127 | -0.0127 | -0.0127 | -0.0128 | -0.0128 | -0.0128 |
| 1.2 microns | -0.0114 | -0.0114 | -0.0114 | -0.0114 | -0.0114 | -0.0114 | -0.0114 | -0.0114 | -0.0114 | -0.0114 | -0.0114 | -0.0115 | -0.0115 |
| 1.5 microns | -0.0118 | -0.0118 | -0.0118 | -0.0118 | -0.0117 | -0.0117 | -0.0117 | -0.0117 | -0.0117 | -0.0118 | -0.0118 | -0.0118 | -0.0118 |
| 1.6 microns | -0.0123 | -0.0123 | -0.0123 | -0.0123 | -0.0123 | -0.0123 | -0.0123 | -0.0123 | -0.0123 | -0.0123 | -0.0123 | -0.0123 | -0.0123 |
| 1.8 microns | -0.0133 | -0.0133 | -0.0133 | -0.0133 | -0.0133 | -0.0133 | -0.0133 | -0.0133 | -0.0133 | -0.0133 | -0.0133 | -0.0133 | -0.0133 |
| 2.0 microns | -0.0147 | -0.0146 | -0.0146 | -0.0146 | -0.0146 | -0.0146 | -0.0146 | -0.0146 | -0.0146 | -0.0146 | -0.0147 | -0.0147 | -0.0147 |
| 2.2 microns | -0.0162 | -0.0162 | -0.0162 | -0.0162 | -0.0162 | -0.0162 | -0.0162 | -0.0162 | -0.0162 | -0.0162 | -0.0162 | -0.0162 | -0.0162 |
| 2.4 microns | -0.0178 | -0.0178 | -0.0178 | -0.0178 | -0.0178 | -0.0178 | -0.0178 | -0.0178 | -0.0178 | -0.0178 | -0.0177 | -0.0177 | -0.0177 |
| 2.5 microns | -0.0191 | -0.0191 | -0.0191 | -0.0191 | -0.0191 | -0.0191 | -0.0191 | -0.0191 | -0.0191 | -0.0190 | -0.0190 | -0.0190 | -0.0190 |

Table 6 – thermo-optic coefficient, dn/dT, of Corning 7980 fused silica at selected wavelengths and temperatures in units of 1/K

| wavelength | 30 K | 40 K | 50 K | 60 K | 80 K | 100 K | 120 K | 160 K | 200 K | 240 K | 275 K | 295 K | 300 K |
|---|---|---|---|---|---|---|---|---|---|---|---|---|---|
| 0.4 microns | 1.25E-06 | 1.60E-06 | 1.95E-06 | 2.30E-06 | 3.00E-06 | 3.70E-06 | 4.67E-06 | 6.02E-06 | 7.11E-06 | 8.20E-06 | 9.16E-06 | 9.70E-06 | 9.84E-06 |
| 0.5 microns | 1.28E-06 | 1.61E-06 | 1.95E-06 | 2.28E-06 | 2.94E-06 | 3.60E-06 | 4.47E-06 | 5.69E-06 | 6.70E-06 | 7.72E-06 | 8.61E-06 | 9.12E-06 | 9.25E-06 |
| 0.6 microns | 1.22E-06 | 1.55E-06 | 1.88E-06 | 2.20E-06 | 2.86E-06 | 3.51E-06 | 4.35E-06 | 5.52E-06 | 6.51E-06 | 7.50E-06 | 8.36E-06 | 8.86E-06 | 8.98E-06 |
| 0.7 microns | 1.26E-06 | 1.58E-06 | 1.89E-06 | 2.20E-06 | 2.83E-06 | 3.46E-06 | 4.27E-06 | 5.43E-06 | 6.40E-06 | 7.36E-06 | 8.21E-06 | 8.69E-06 | 8.81E-06 |
| 0.8 microns | 1.16E-06 | 1.48E-06 | 1.80E-06 | 2.13E-06 | 2.77E-06 | 3.42E-06 | 4.23E-06 | 5.36E-06 | 6.32E-06 | 7.28E-06 | 8.12E-06 | 8.60E-06 | 8.72E-06 |
| 0.9 microns | 1.17E-06 | 1.49E-06 | 1.81E-06 | 2.13E-06 | 2.76E-06 | 3.40E-06 | 4.21E-06 | 5.33E-06 | 6.27E-06 | 7.22E-06 | 8.05E-06 | 8.52E-06 | 8.64E-06 |
| 1.0 microns | 1.07E-06 | 1.40E-06 | 1.74E-06 | 2.07E-06 | 2.73E-06 | 3.39E-06 | 4.21E-06 | 5.30E-06 | 6.24E-06 | 7.18E-06 | 8.01E-06 | 8.48E-06 | 8.60E-06 |
| 1.2 microns | 8.86E-07 | 1.26E-06 | 1.64E-06 | 2.01E-06 | 2.76E-06 | 3.52E-06 | 4.29E-06 | 5.25E-06 | 6.19E-06 | 7.12E-06 | 7.94E-06 | 8.40E-06 | 8.52E-06 |
| 1.5 microns | 9.92E-07 | 1.34E-06 | 1.69E-06 | 2.04E-06 | 2.73E-06 | 3.43E-06 | 4.19E-06 | 5.22E-06 | 6.18E-06 | 7.13E-06 | 7.97E-06 | 8.45E-06 | 8.57E-06 |
| 1.6 microns | 9.31E-07 | 1.30E-06 | 1.66E-06 | 2.02E-06 | 2.75E-06 | 3.48E-06 | 4.27E-06 | 5.23E-06 | 6.14E-06 | 7.05E-06 | 7.84E-06 | 8.29E-06 | 8.41E-06 |
| 1.8 microns | 9.16E-07 | 1.27E-06 | 1.62E-06 | 1.98E-06 | 2.68E-06 | 3.39E-06 | 4.19E-06 | 5.22E-06 | 6.16E-06 | 7.09E-06 | 7.91E-06 | 8.38E-06 | 8.50E-06 |
| 2.0 microns | 8.22E-07 | 1.17E-06 | 1.51E-06 | 1.85E-06 | 2.54E-06 | 3.22E-06 | 4.03E-06 | 5.14E-06 | 6.13E-06 | 7.12E-06 | 7.98E-06 | 8.48E-06 | 8.60E-06 |
| 2.2 microns | 1.01E-06 | 1.37E-06 | 1.73E-06 | 2.09E-06 | 2.82E-06 | 3.54E-06 | 4.35E-06 | 5.28E-06 | 6.12E-06 | 6.96E-06 | 7.69E-06 | 8.11E-06 | 8.22E-06 |
| 2.4 microns | 1.14E-06 | 1.45E-06 | 1.77E-06 | 2.08E-06 | 2.71E-06 | 3.34E-06 | 4.12E-06 | 5.21E-06 | 6.15E-06 | 7.08E-06 | 7.91E-06 | 8.38E-06 | 8.49E-06 |
| 2.5 microns | 1.06E-06 | 1.40E-06 | 1.74E-06 | 2.08E-06 | 2.77E-06 | 3.46E-06 | 4.08E-06 | 5.10E-06 | 6.17E-06 | 7.24E-06 | 8.18E-06 | 8.71E-06 | 8.85E-06 |
| 2.6 microns | 1.03E-06 | 1.38E-06 | 1.72E-06 | 2.07E-06 | 2.76E-06 | 3.45E-06 | 4.22E-06 | 5.25E-06 | 6.20E-06 | 7.16E-06 | 8.00E-06 | 8.48E-06 | 8.60E-06 |

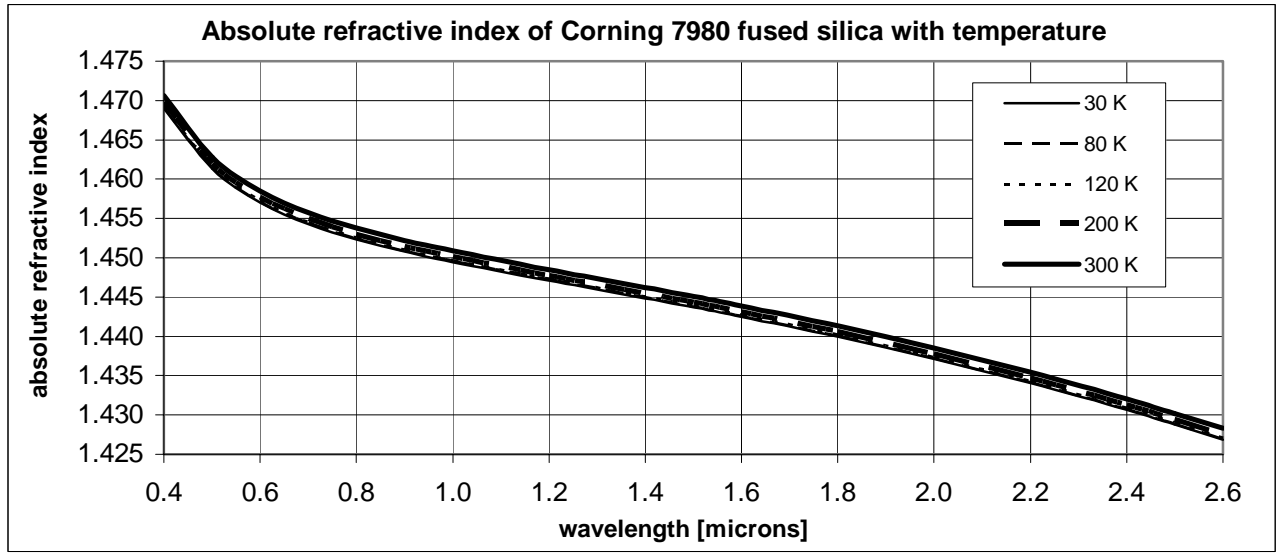
Figure 4 – measured absolute refractive index, n, of Corning 7980 fused silica at selected wavelengths and temperatures

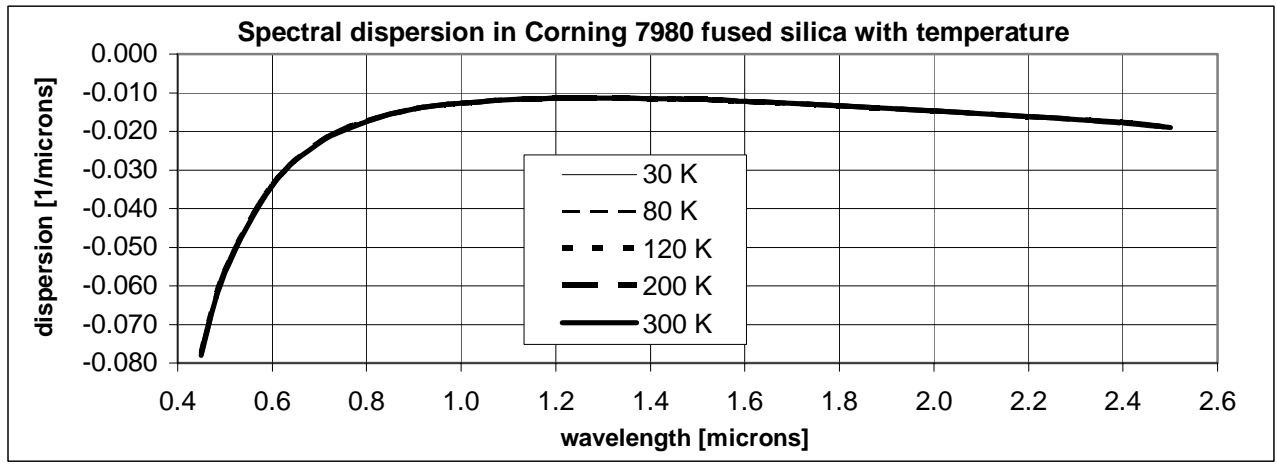
Figure 5 – spectral dispersion, dn/dλ, in Corning 7980 fused silica at selected temperatures

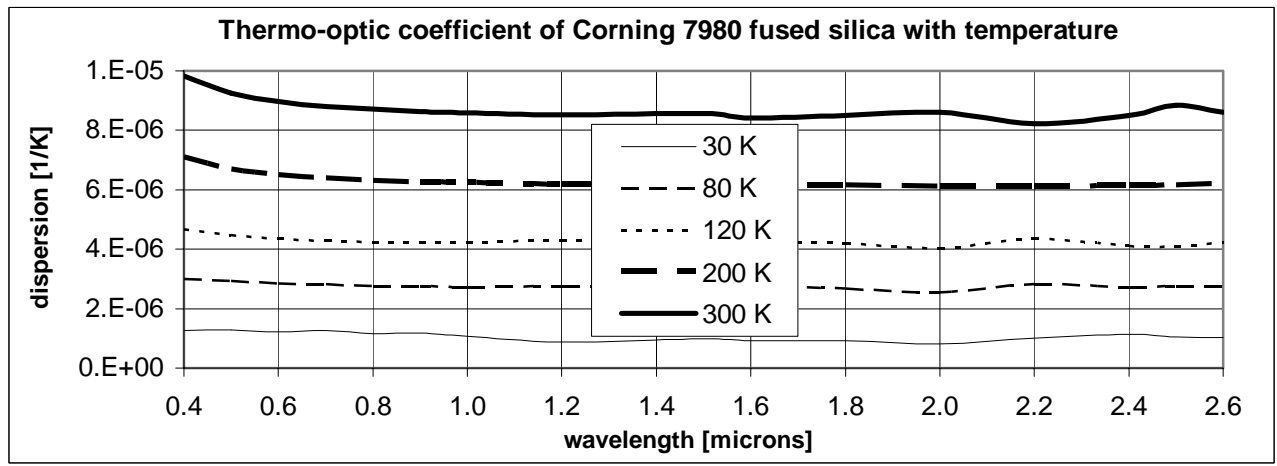
Figure 6 – thermo-optic coefficient, dn/dT, of Corning 7980 fused silica at selected temperatures

# 4. COMPARISON WITH PREVIOUS INVESTIGATIONS

In this section, we compare the results of our absolute refractive index study on fused silica with those listed in the second paragraph of Section 1. Figure 7 illustrates the wavelength and temperature coverage of the various studies. Not shown in the chart are studies which extended to wavelengths shortward of the near ultraviolet near room temperature conducted by a number of investigators including ourselves.

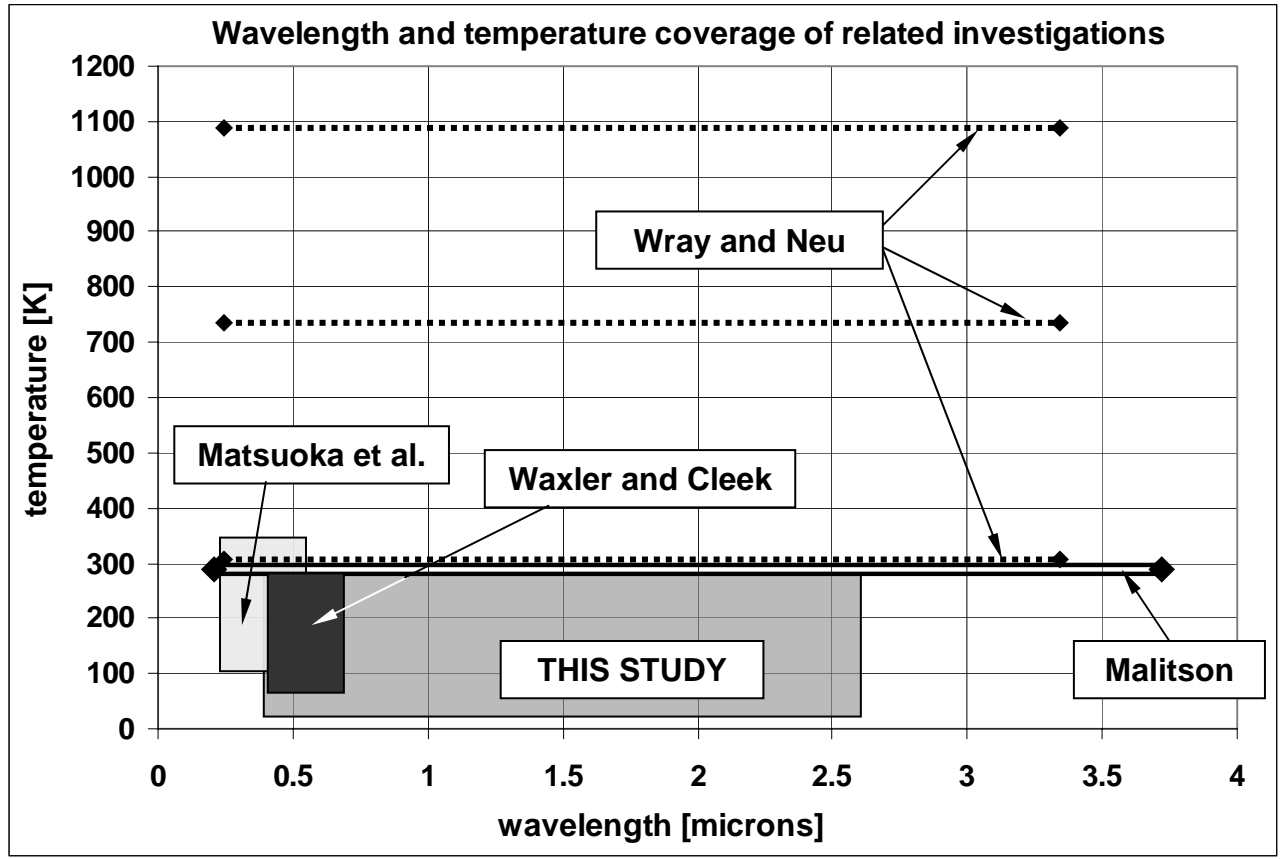

Figure 7 – wavelength and temperature coverage of this study and of other investigations on the refractive index of fused silica

We have already compared our results to Malitson's results in Section 2. The measurements of Wray and Neu are outside the temperature range considered in this study but have been put on this map to show their context to those of other investigators. Fortunately, we are in a position to compare our results to those of Waxler and Cleek (W&C) and to those of Matsuoka et al. who did perform measurements at cryogenic temperatures albeit by different means. Remember that W&C used an interferometric technique – similar in principle to that used by PMIC for the Kepler program – to measure only changes in refractive index with respect to room temperature appealing to room temperature data of Malitson (which are in air) and to CTE values for fused silica from other investigators to compute values of index at temperatures within their measurement range. Although our measured wavelength range overlaps only from mid-visible through violet wavelengths with that of Matsuoka et al., the latter measurements were made using essentially the same technique we use – the minimum deviation method – which does not appeal to external databases on physical properties, and so we feel make for a more valid and interesting comparison.

Table 7 compares W&C's index measurements to the measurements in CHARMS. In order to do this comparison, we first calculate refractive index using our Sellmeier model at the wavelengths and temperatures used by W&C. We then correct W&C's measurements to be in vacuum, and then take the difference between adjusted W&C and CHARMS indices. Near room temperature, there is good agreement as expected since W&C used Malitson's data to anchor their indices at room temperature. However, curiously W&C's data are lower than those from CHARMS by several parts in the fifth decimal place of index at temperatures below 160 K, just as were those from PMIC who used a similar measurement technique and manipulation of CTE data to derive refractive index.

Table 8 compares absolute refractive index measurements of Matsuoka et al. to measurements in CHARMS. Matsuoka's indices agree with indices from CHARMS to within about twice our stated uncertainty of 0.00001. CHARMS indices appear to be systematically higher than those of Matsuoka by 1 to 2 parts in the fifth decimal place of index. This is most likely explainable through differences in source of supply of the materials measured.

Table 7 – adjusted refractive indices of Waxler and Cleek for fused silica minus refractive indices from CHARMS

| wavelength [microns] | 73 K | 123 K | 173 K | 223 K | 273 K | 293 K |
| --- | --- | --- | --- | --- | --- | --- |
| 0.668 | -0.00004 | -0.00006 | -0.00006 | -0.00002 | 0.00000 | 0.00000 |
| 0.644 | -0.00001 | -0.00008 | -0.00005 | 0.00000 | 0.00001 | -0.00001 |
| 0.588 | -0.00004 | -0.00006 | -0.00005 | -0.00001 | 0.00000 | 0.00000 |
| 0.509 | -0.00005 | -0.00006 | -0.00004 | -0.00001 | 0.00000 | 0.00000 |
| 0.502 | -0.00008 | -0.00006 | -0.00004 | -0.00001 | 0.00000 | 0.00001 |
| 0.480 | -0.00004 | -0.00006 | -0.00004 | -0.00002 | 0.00000 | 0.00000 |
| 0.471 | -0.00005 | -0.00006 | -0.00003 | 0.00000 | 0.00001 | 0.00001 |
| 0.468 | -0.00003 | -0.00006 | -0.00003 | 0.00000 | 0.00001 | 0.00001 |
| 0.436 | -0.00005 | -0.00003 | -0.00001 | 0.00003 | 0.00003 | 0.00001 |
| 0.405 | -0.00002 | -0.00005 | -0.00003 | 0.00001 | 0.00002 | 0.00002 |

Table 8 – absolute refractive indices of Matsuoka et al. for fused silica minus refractive indices from CHARMS

| wavelength [microns] | 108 K | 162 K | 230 K | 276 K | 294 K |
| --- | --- | --- | --- | --- | --- |
| 0.546 | -0.00002 | -0.00002 | -0.00002 | -0.00002 | -0.00002 |
| 0.436 | -0.00002 | -0.00002 | -0.00001 | -0.00001 | -0.00001 |
| 0.405 | -0.00001 | -0.00002 | -0.00001 | 0.00000 | 0.00000 |

## 6. SUMMARY

Using CHARMS, we made over 7600 individual, direct measurements of absolute refractive indices of modern synthetic fused silica (Corning 7980) from 30 to 300 K to an accuracy level of 1 part in the fifth decimal place of index across the visible spectrum and through the infrared out to the strong absorption in the material at 2.75 microns wavelength, greatly expanding the wavelength and temperature coverage for measurements of this material, especially to cryogenic temperatures. Spectral dispersion appears to be a very weak function of temperature. We demonstrated that five different specimens taken from around the perimeter of a 1 m diameter optical blank for the Schmidt telescope corrector for the Kepler Photometer share the same index to within our ability to measure it at all temperatures measured. Our cryogenic measurements compare favorably to sparse measurements from previous investigations.

## ACKNOWLEDGEMENTS

The authors wish to Charles Sobeck at NASA/Ames and Don Byrd/Ball Aerospace Corporation for providing the test samples and the comparison dn/dT data from the sister investigation at PMIC to enhance the completeness of this study.